# Tunable spontaneous emission from layered graphene/dielectric tunnel junctions

Sina Khorasani
School of Electrical Engineering, Sharif University of Technology, P. O. Box 11365-9363, Tehran, Iran
e-mail: khorasani@sharif.edu

There has been a rapidly growing interest in optoelectronic properties of graphene and associated structures. Despite the general belief on absence of spontaneous emission in graphene, which is normally attributed to its unique ultrafast carrier momentum relaxation mechanisms, there exist a few recent evidences of strong optical gain and spontaneous light emission from mono-layer graphene, supported by observations of dominant role of out-of-plane excitons in polycyclic aromatic hydrocarbons. In this article, we develop a novel concept of light emission and optical gain from simple vertical graphene/ dielectric tunnel junctions. It is theoretically shown that the possible optical gain or emission spectrum will be easily tunable by the applied voltage. We present details of quantum mechanical calculations and perform an exact analysis of field-assisted tunneling using transfer matrices combined with expansion on Airy's functions.

## I. Introduction

Electro-luminescence is the primary method for energy efficient conversion of electricity into light. In all semiconductor based materials this happens as a result of radiative recombination of electrons and holes. For this purpose, semiconductors with direct bandgaps are used. Meanwhile, quantum confinement of carriers by artificial nanostructures helps to amplify the electroluminescence through increasing the dipole moments by two orders of magnitude, thereby obtaining bright light sources in the optical spectrum, ranging from near infrared to the ultraviolet. In all cases, the choice of color is fixed once the material is chosen and the structure is fabricated. Application of external voltage helps the device to change its brightness. Hence, a red-emitting device, for instance, can never emit in the blue spectrum. That means the emission spectra are more or less fixed at the time of fabrication.

The peculiar electronic properties of graphene, the two-dimensional wonder material, here come to help. This zero-bandgap semimetal exhibits a linear dispersion around the so-called Dirac points, allowing color tunability through careful design. While it had been believed that ultrafast mechanisms would prevent light emission [1,2], ultrafast electro-luminescence [3], optical gain [4], and population inversion and THz lasing [5] have recently been directly observed in graphene. A simple calculation reveals that graphene permits very large dipole moment in the optical spectrum, indeed, quite comparable to those of the best engineered quantum confined nanostructures.

In the proposed design, two graphene layers are separated by an extremely thin dielectric, which should be just a few Angstroms in width. The best choice for this purpose might be hexagonal boron nitride, the hydrogenated graphene (graphane), or graphene-oxide. Graphene layers on both sides are pristine and can be un-doped. The planar nature of dielectric allows low-voltage operation and high brightness, because of the peculiar form of the density of states. Such heterojunctions have been in use [6-10] for study of transistor actions and photonic modulators. Alternative materials are oxides and sulfides of tungsten and molybdenum [10-12]. Recently, photoluminescence has been observed in mono-layer tungsten disulfide [13,14], graphane [15-18], and transition metal dichalcogenides [19]. With the advent of graphene-like planar materials such as graphane [15-19], silicene [20], and germanane [21] even further possibilities could be imagined. It should be also mentioned that a comparable tunnel structure made of p-Si/Thin $SiO_2$/n-Si [22] has been shown to enhance the infrared light emission by up to 3 orders of magnitude.

Application of an external voltage causes a shift in the Fermi levels of both sides, but in opposite directions, so that the cathode (anode) gets occupied (depleted) of high (low) energy electrons. The amount of difference in the energies on the two sides is roughly proportional to the square root of the applied voltage, when un-doped graphene layers are used for both electrodes. This is in contrast to the in-plane excitons which normally live in single-layer materials, where recombinations of out-of-plane excitons are here sought [23]; interestingly, it has been also shown that out-of-plane excitons are just as important as in-plane excitons in light emission from polycyclic aromatic hydrocarbons, such as graphene, pentacene, and perylene [23]. Now, high energy electrons can tunnel across the barrier through the field-assisted tunneling mechanism, and undergo recombination with holes on the anode surface. The maximum in the emission spectrum is therefore equal to the twice value of the Fermi level, which is in turn proportional to the square root of the voltage. As a result, the output spectrum gets blue-shifted by increasing the voltage, and color tunability is obtained simply by changing the voltage.

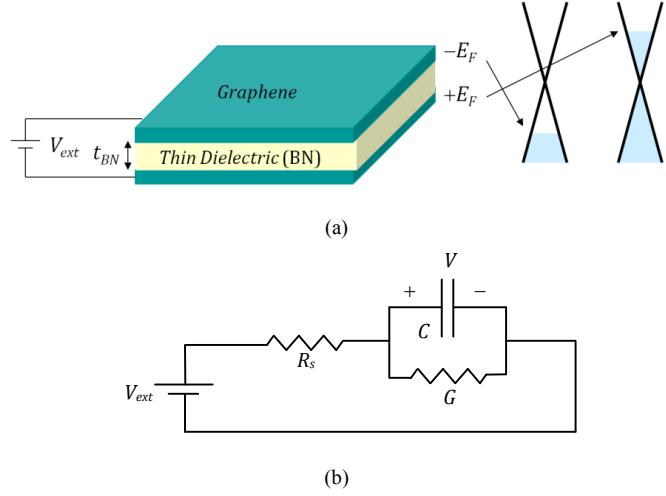

Fig. 1. Proposed graphene/boron-nitride tunnel junction subject to an applied external voltage: (a) structure; (b) equivalent circuit: $R_s$ is a series resistance representing the series Ohmic voltage drops on the contacts, $G$ is the quantum tunneling conductance, and $C$ is the capacitance of the structure.

A detailed computation reveals that all visible colors are accessible by voltages ranging from 0.35 to 0.8 volts. Multiple layers can be added to obtain higher brightness, evidently if emission rate is sufficiently higher than the absorption (for a spontaneous light emitting device this criterion is normally met at the expense of some additional input power). Meanwhile for large-area structures, electrodes can be patterned to obtain an addressable and compact dot-matrix of light emitting diodes. Raster scanning will therefore enable production of live videos.

As illustrated in Fig. 1, this structure is at the same time mechanically transportable and very much flexible. It can be used virtually everywhere, ranging from books covered with electro-optic illustrations, cloths, low-cost electro-optic advertisements, etc., simply implying an enabling technology to have displays everywhere on the planet.

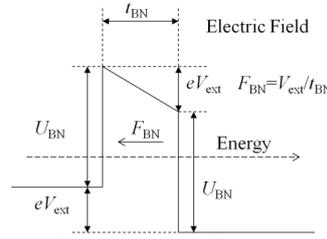

Fig. 2. Energy diagram illustrating the field-assisted tunneling across the graphene/boron-nitride junction.

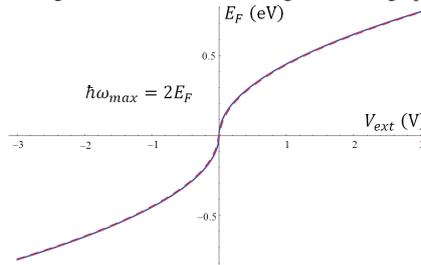

Fig. 3. Plot of relationship between external applied voltage and shift of Fermi levels: approximate (red-dashed) and exact (solid-blue).

## II. Structure and Calculations

The tunneling mechanism in the graphene tunnel junction of interest can be illustrated as shown in Fig. 2. Here, regions I and III represent graphene sheets while region II is the barrier dielectric. If taken a mono-layer of boron-nitride, the corresponding thickness and barrier height will be respectively, $T_{BN} = 2.489\text{Å}$, and $U_{BN} = \chi_{h_{BN}} - \frac{1}{2}E_{g_{BN}} = 3.28\text{eV}$, with hole affinity $\chi_{h_{BN}} = 5.8\text{eV}$ and energy band gap $E_{g_{BN}} = 5.05\text{eV}$. Application of an external voltage $V_{ext}$ would cause a shift between the Fermi levels as well as establishing an assisting electric field in the barrier. Symmetry of voltage distribution is a simple result of the facts that (a) the densities of accumulated holes and electrons must be equal on both sides because of charge conversation; (b) the band-structure of graphene is nearly symmetric for holes and electrons across the Dirac point. The exact tunneling probability is calculated as follows later.

The amount of shift of Fermi levels on each side is thus evidently symmetric, which may be estimated using the red-dashed square-root relationship as shown in Fig. 3, given by the square-root dependence $E_F \propto \text{sgn}(V)\sqrt{|V|}$, where the proportionality constant may be determined from integration over Fermi-Dirac distribution multiplied by the density of states for the two-dimensional graphene (in the rest of this paper, the symbol $E_F > 0$ is reserved for the electron Fermi level). This will result in the approximate relationship (red-dashed curve in Fig. 3)

$$E_F = \left(\pi \hbar v_F \sqrt{\frac{\epsilon_{BN}}{q t_{BN}}}\right) \text{sgn}(V)\sqrt{|V|} \quad (1)$$

in which $\epsilon_{BN} = 4\epsilon_0$ is the dielectric constant of the barrier material and $q$ being the electronic charge. The square-root dependence (1) has been also in use for similar structures [24], and is to be compared versus the exact relationship (solid-blue curve in Fig. 3) obtained from direct integration over the electronic density of states, times the corresponding Fermi-Dirac distributions.

It has to be here mentioned that in practice metal contacts need to be placed somewhere on the graphene layers (not shown in Fig. 1 above), causing a shift $\Delta V_{ext}$ in $V_{ext}$ due to the corresponding Schottkey-barrier voltage drops (interestingly, a novel method has been also just reported [25], which enables high-quality one-dimensional contact to graphene across its edge, instead of depositing metal contacts on the top of graphene surface). Hence, it is the driving voltage $V_{ext} = V + \Delta V_{ext}$ that is actually applied externally to the combined structure (metal contacts, graphenes, and dielectric layers). All these effects may be combined into a nonlinear series resistance $R_s$ as illustrated in the equivalent circuit in Fig. 1b. Evidently, if symmetric metal contacts with work functions close to that of the graphene are to be used, then the overall voltage difference would roughly smooth out, as one junction is forward-biased while the other is reverse-biased. Hence, for the sake of brevity we may take the Ohmic linear approximation $\Delta V_{ext} \approx R_s I$ and disregard such small non-linear terms. The DC current flowing into the structure will pass through the quantum tunneling conductance $G = A G_g$ to be calculated later, placed in parallel to the capacitance $C = A C_g$. Here, $A$ is the area of the device, $C_g = \epsilon_{BN}/t_{BN}$, and $G_g$ is the per-unit quantum conductance to be found later. It may be added here that a three-terminal closely related device has been already proposed as a tunnel-FET [26].

While the two-dimensional density of states for gapped materials is given by [27] $\varrho_G(E) \propto |E|^{\frac{1}{2}(N-2)}$ with $N = 2$ being the dimensionality, for the gapless graphene we have $\varrho_G(E) \propto |E|$ [28]. Eventually, the Fermi energy $E_F$ may be determined by solution of the implicit relationship

$$2\left(\frac{k_B T}{\pi \hbar v_F}\right)^2 \left[-\text{Li}_2\left(-e^{+\frac{|E_F|}{k_B T}}\right) + \text{Li}_2\left(-e^{-\frac{|E_F|}{k_B T}}\right)\right] = \frac{\epsilon_{BN}|V|}{q t_{BN}} \quad (2a)$$

$$-\text{Li}_2(-e^{+|E_F|/k_B T}) = \frac{1}{k_B T}\int_0^\infty f_e[E] \frac{E}{k_B T} dE \quad (2b)$$

$$-\text{Li}_2(-e^{-|E_F|/k_B T}) = \frac{1}{k_B T}\int_{-\infty}^0 f_h[E] \frac{-E}{k_B T} dE \quad (2c)$$

in which $k_B T$ is the thermal energy obtained by multiplication of the Boltzmann's constant $k_B$ and absolute temperature $T$, $\text{Li}_n(x)$ is the poly-Log function, and electron and hole Fermi-Dirac functions $f_e[E]$ and $f_h[E]$ are shown below in (4). Accordingly, the Fermi levels across the graphene/ insulator/graphene system is adjustable such that $E_F \propto \sqrt{|V|}$. Later we shall see that this value also roughly corresponds to the maximum of emission spectra according to $\hbar \omega_{max} = 2 E_F$. It should be added that (1) is actually the zero-temperature limit of (2) with $T \rightarrow 0^+$. Solution of (2a) is possible only numerically, which is to be compared with the approximate relationship (1) in Fig. 3.

Ignoring the non-radiative recombination mechanisms (see Appendix A), the intensity of emitted photons at a given wavelength can be estimated by considering the tunneling probability of electrons or holes through the insulator, electron and holes distributions over energy, and transition matrix elements. Mathematically one can consider all transitions over the Brillouin zone and write the luminance function (having the dimension of energy density) as

$$L_v(\omega) == \frac{\hbar \omega}{q^2 S_{UC}} \int_{BZ} |M(\mathbf{k}, \omega)|^2 |T[E_\mathbf{k}]|^2 f_e[E_\mathbf{k}] f_h[-E_\mathbf{k}] L(\mathbf{k}, \omega) \varrho(\mathbf{k}) d^2 k \quad (3)$$

Here, $S_{UC} = 3\sqrt{3}a^2/2$ with $a = 0.142$nm being the carbon-carbon bond length is the area of unit cell in graphene, $L(\mathbf{k},\omega)$ accounts for a Lorentzian broadening, $M(\mathbf{k},\omega)$ is the transition matrix element for electron-hole recombination. $E_\mathbf{k} = \hbar v_F k$ represents the conical band-structure of graphene expanded at its Dirac points, in which $v_F$ is the Fermi velocity. $|T[E_\mathbf{k}]|^2$ is tunneling probability, and $f_e[E_\mathbf{k}]$ and $f_h[-E_\mathbf{k}]$ are the electron and hole distributions on opposite sides, given by

$$f_e[E_\mathbf{k}] = \{1 + \exp[(E_\mathbf{k} - E_F)/k_B T]\}^{-1}$$
$$f_h[-E_\mathbf{k}] = 1 - \{1 + \exp[(-E_\mathbf{k} + E_F)/k_B T]\}^{-1}$$
(4)

Also, $\varrho(\mathbf{k})$ is the graphene's density of states in the wavevector space, given by

$$\varrho(\mathbf{k})d^2k = 6\frac{1}{\pi^2}d^2k = 6\frac{2|E|}{\pi(\hbar v_F)^2}dE = \varrho(E)dE$$
(5)

Here, we notice that the Fermi level for holes $-E_F$ is a negative quantity. Also, the transition frequency $\hbar\omega_\mathbf{k} = 2E_\mathbf{k}$ corresponds to the frequency of the photon resulting from the direct band-to-band radiative recombination of an electron-hole pair with identical momenta of $k = |\mathbf{k} - \mathbf{k}_F|$. The factor 6 is the number of Dirac points centered at various $\mathbf{k}_F$'s.

Connected with any tunneling/transition we may expect broadenings as results of thermal processes as well as uncertainties in energies and momenta. The combined effects may be appropriately modeled by

$$L(\mathbf{k},\omega) = \frac{\hbar}{\tau\pi}\left\{[2E_\mathbf{k} - \hbar\omega]^2 + \left(\frac{\hbar}{\tau}\right)^2\right\}^{-1}$$
(6)

where $\tau$ is the associated phenomenological time constant. We anticipate this parameter to be of the order of 1ps corresponding to an uncertainty in energy of the order of 1meV. Larger values than 10ps effectively have no appreciable effect on the final spectrum, while shorter durations result in strong diminishing of the output optical power. The same order also applies to similar optoelectronic processes in direct-gap semiconductors, nevertheless, a reliable estimation needs accurate experiments.

The transition matrix elements of graphene are given by [29]

$$\langle c_\mathbf{k}|\hat{e}\cdot\hat{v}|v_\mathbf{k}\rangle = \frac{v_F}{k}|k_x e_y - k_y e_x|$$
(7)

in which $\hat{e} = e_x\hat{x} + e_y\hat{y}$ is the unit polarization vector and $\hat{v}$ is the velocity operator. For the case of spontaneously emitted single photons from a radiative recombination, only circular polarizations may be expected. This would simplify (7) as $|\langle c_\mathbf{k}|\hat{e}\cdot\hat{v}|v_\mathbf{k}\rangle| = v_F$. Now, the electric dipole moments are $M(k,\omega) = q\langle c_\mathbf{k}|\hat{e}\cdot\hat{v}|v_\mathbf{k}\rangle$, which can be obtained by noting the fact that $\hat{v} = \frac{i}{\hbar}[\mathbb{H},\hat{r}]$ where $\mathbb{H}$ is the Hamiltonian. After some algebra (Appendix B) we are able to get

$$|M(\mathbf{k},\omega)| = qv_F/\omega = 10^{21}\,q\lambda v_F/2\pi\ \text{(Debye)}$$
(8)

This result shows that in the visible and infrared spectrum, pristine Graphene is nearly as good as the best quantum confined nanostructures in regard to their very large transition dipoles as illustrated in Fig. 4, being typically in the range of 40-80 Debye.

It is known that the transition dipoles significantly increase for quantum confined structures compared to the bulk materials. Comparison of the transition dipoles for those of 2D quantum confined heterostructures can be easily made. At the wavelength of 1.55μm the values reported in [31] are 26.15 Debye and 15.21 Debye respectively for e-hh and e-lh transitions in a lattice-matched InGaAlAs/InP quantum well, while the same value for the pristine (unconfined and unpatternd) graphene at the same wavelength is 39.47 Debye. At larger wavelengths, the transition dipole of graphene continues to increase linearly, while for semiconductors it remains limited to within the same order of magnitude.

At 1THz the dipole of graphene reaches the value of 7400 Debye, which explains the unparalleled performance of this material at millimeter waves. This peculiar property of Graphene is owing to its unique band-structure and existence of Dirac points, as well as small C-C bond length and hence large orbital overlaps between $\pi$ and $\pi^*$ bands.

TABLE I
Shift in Operating Voltages for Various Cathode Materials

|  | Al | Au | Fe | Cu | Ag |
|---|---|---|---|---|---|
| $V_{ext}(V)$ | -0.42 | 0.6 | 0 | 0.20 | 0.22 |

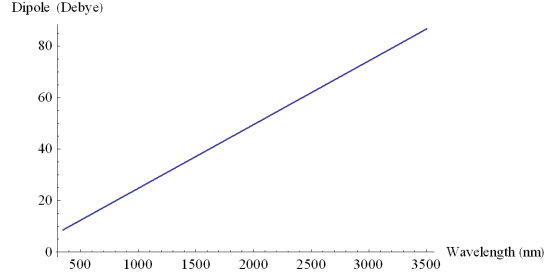

Fig. 4. Calculated transition dipole of pristine graphene versus wavelength.

The tunneling probability $|T[E]|^2$ of the electrons through the insulator according to Fig. 2, depends not only on the thickness of the potential barrier but also on the applied electric field. A detailed treatment of the problem of finding the field-assisted tunneling probability in such a system using transfer matrix formalism [30,31] combined with expansion on Airy's functions [32] is briefly explained here.

The electron wavefunctions in regions I and III are given by $\psi_j(z) = a_j^+ e^{iK_j z} + a_j^- e^{-iK_j z}, j = I, III$, with wavenumbers $K_I = \sqrt{2mE/\hbar^2}$ and $K_{III} = \sqrt{2m(E+qV)/\hbar^2}$. However, in region II we have $\psi_{II}(z) = a_{II}^+ \text{Ai}[K_{II}(z-z_0)] + a_{II}^- \text{Bi}[K_{II}(z-z_0)]$ based on the Airy's functions of the first and second type, $K_{II} = \sqrt[3]{-2mqF/\hbar^2}$, $z_0 = (U_{BN}-E)/qF$, and $F = V_{ext}/T_{BN}$. By applying the continuity of wavefunctions and their derivatives at $z=0$ and $z=T_{BN}$ we obtain four equations in terms of the six unknowns $a_j^\pm, j = I, II, III$. These can be solved by setting $a_I^+ = 1$ and $a_{III}^- = 0$ to obtain the tunneling probability $|T|^2 = |a_{III}^+|^2$. Analytical form does exist fortunately and may be obtained after significant algebra, which takes the form

$$T = \frac{2K_{II} K_I}{\pi G}$$
$$G = (K_{II}A + iK_{III}a)(K_I b - iK_{II} B) + (iK_I c + K_{II} C)(K_{III} d + iK_{II} D)$$
$$a = \text{Ai}(\alpha), A = \text{Ai}'(\alpha), c = \text{Ai}(\beta), C = \text{Ai}'(\beta)$$
$$b = \text{Bi}(\beta), B = \text{Bi}'(\beta), d = \text{Bi}(\alpha), D = \text{Bi}'(\alpha)$$
$$\alpha = -\frac{2m(E + qV - U_{BN})}{\hbar^2 K_{II}^2}$$
$$\beta = -\frac{2m(E - U_{BN})}{\hbar^2 K_{II}^2}$$

(9)

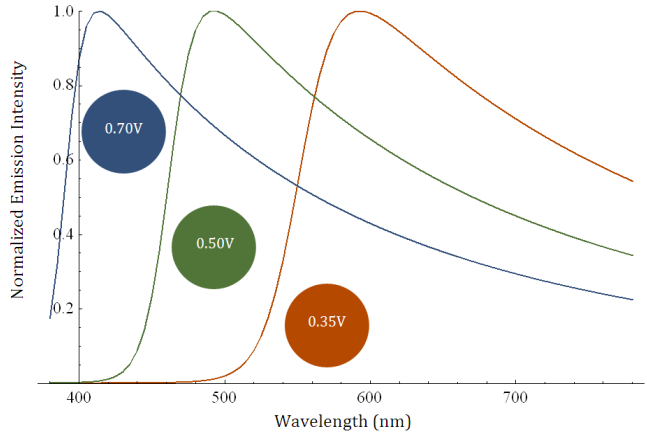

Fig. 5. Calculated spectrums of emitted spontaneous light at various applied voltages with expected apparent colors found from standard CIE indices.

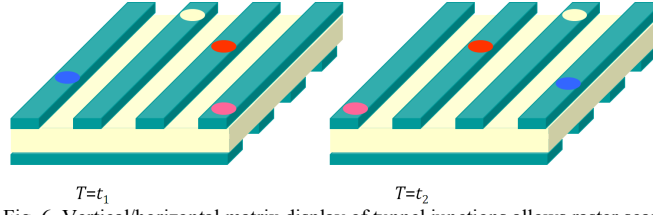

Fig. 6. Vertical/horizontal matrix display of tunnel junctions allows raster scan.

In Fig. 5 the normalized emission spectra of the system under a few various different bias voltages are plotted for circularly polarized emitted spontaneous light. Clearly the peak wavelength of the emission spectra is a function of external voltage. As it was expected, we may easily observe the drift of emission maxima with the applied external voltage according to $\lambda_{max} = 2\pi\hbar c/q|V_{ext}|$.

This suggests that the color of the emitted light perceived by human eye can be adjusted and tuned in a rather continuous way by applying the external voltage. According to the plot, the FWHM at the blue end of the emission spectrum is around 200nm, while it increases for longer wavelengths at the red end of the spectrum to above 240nm. Evidently, the spectra are asymmetric around their maxima due to the sharp decrease of Fermi-Dirac distribution above the Fermi energy. Of course, this device is not able to emit any desired color especially because of its relatively wide FWHM, and for this purpose it might not be appropriate for stimulated emission applications. However, combinations of various colors via three pixels may produce any desired color code. For this purpose, any pixel must be consisting of three of such emitters, and hence a four-terminal device at least. In Fig. 5 the emission spectra are converted to their corresponding colors based on CIE standard [33].

One of the interesting possible applications of this light emitting junction is the realization of a matrix array of light emitting cells, which at each crossing would emit a certain color controllable by the momentary applied voltage to the two electrodes. Raster scanning will allow fully functional live videos over extremely thin and flexible light emitting thin film displays, as shown in Fig. 6.

Counter-intuitively, the photon energies in the presented design exceed the equivalent energy due to the external bias. This contradiction is in fact due to the square-root dependence of Fermi-energy $E_F$ on the external voltage $V_{ext}$ (1), and could be explained as follows. Let for positive voltages we have $E_F = D\sqrt{qV} \approx D\sqrt{qV_{ext}}$ where $D$ is the proportionality constant. This approximatuion is true when the voltage drop on the series contact resistance $R_s$ is negligible, which is the case for small current densities. Evidently, for $D > \frac{1}{2}\sqrt{qV}$, or external voltages satisfying $V < V_{ext} < V_M = 4D^2/q$, we obtain $2E_F > qV_{ext}$. Hence, it is not surprising that the Fermi energy could exceed the apparent bias energy and violate the expected relationship $qV_{ext} > 2E_F$. The physical reason behind this argument is that the strange density of states function for the gapless graphene is proportional to the energy, whereas for the conventional two-dimensional gapped materials is independent of the energy. Hence, the relationship between the Fermi energy and external voltage in not trivial. For sufficiently small external bias when $V_{ext} < V_M$, and sufficiently small thickness (which is inversely proportional to $D$), the Fermi energy could exceed the bias, or somehow virtually amplified. This hidden boost in the bias voltage is nevertheless irrelevant to the conservation of energy, where output optical power should match the input electrical power including losses. The numerical value of $V_M$ for this structure is calculated to be around 14V.

Now, in contrast to the comparable structure in [24], the main difference is the thickness of the dielectric barrier between the two graphene layers. In our design the BN layer is only one mono-layer thick (to enable large tunneling current), while the work in [24] is 7nm thick, being about 28 times thicker than that of ours. It is not difficult to verify that larger thickness would result in larger drive voltage, while keeping the Fermi levels at the same. The approximate relationship between the Fermi energy and applied voltage according to the article [24] is $E_F = \hbar v_F \sqrt{\eta\pi[V + V_0]}$, where $\eta = \pi \epsilon_{BN}/qt_{BN}$ is proportional to the capacitance per unit area over electronic charge; by assuming $\epsilon_{BN} = 4\epsilon_0$, we obtain $\eta = 9.9 \times 10^{16} m^{-2}V^{-1}$, in agreement with the experimentally fitted value of $\eta = 9 \times 10^{16} m^{-2}V^{-1}$ in the paper [24]. This fit also perfectly matches to the relation (1) above. Now, for our structure with one mono-layer dielectric, the typical operating voltage must be roughly 5 times smaller. Clearly, a thinner dielectric and hence larger capacitance per unit area would require much lower voltage, while keeping the Fermi level unchanged. This would explain the low-voltage operation of our proposed device.

We furthermore may also cite the reference [22], where the fabricated tunneling device emits at 1.14μm, which is equivalent to 1.1eV. The device has been biased well under 1V, while the device area has been 1.5×1.5mm². Hence, the average current density has been less than 4.44 A/cm², corresponding to the bias current of 100mA. The considered p-Si/Ultrathin SiO$_2$/n-Si device [22], has had some appreciable emission around 1.5-3μW for current densities ranging 2-5 A/cm². This range of current density points to a bias voltage well under 1V, where the bias current should have been around 45-110mA. Hence, this device has been capable of emitting 1.5-3μW of optical power at 1.1eV, while the bias voltage has been 0.92-1.0V, well below the 1.1V. Therefore emission of a photon with energy $h\nu$ does not necessarily require a drive voltage larger than $h\nu/q$.

The tunneling current is given by the Landauer formula [34] across the structure

$$J(V) = VG_g = J^+(V) - J^-(V)$$

(10a)

$$J^+(V) = G_0 \int |T[E]|^2 f_e[E; E_F(V)] f_h[-E; E_F(V)] \varrho(E) dE \tag{10b}$$

$$J^-(V) = G_0 \int |T[E]|^2 f_e[-E; -E_F(V)] f_h[E; -E_F(V)] \varrho(E) dE \tag{10c}$$

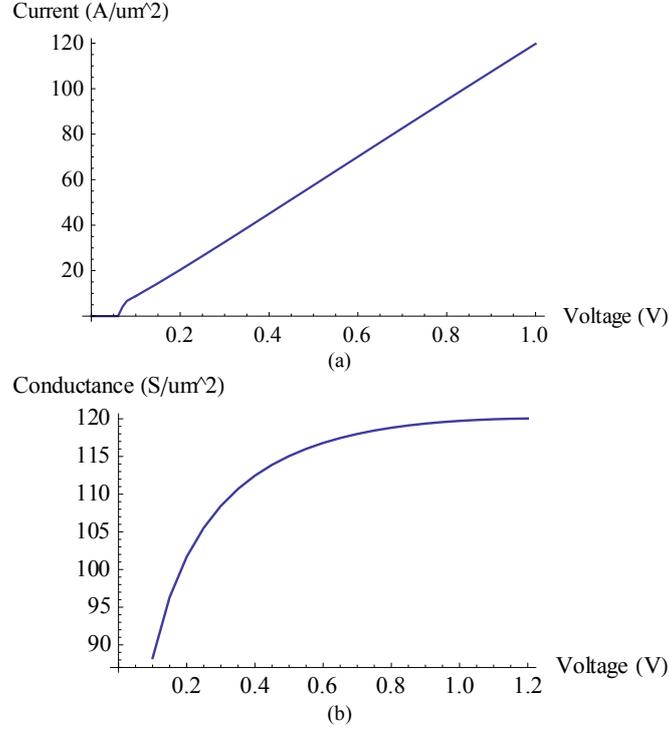

Fig. 7. (a) Nonlinear tunneling current versus applied voltage; (b) The corresponding conductance density.

where $G_0 = q^2/\pi\hbar$ is the quantum conductance. Hence, $G_g$ is the nonlinear quantum tunneling conductance per unit area given by $G_g = J(V)/V$. The I-V relationship and nonlinear conductance are calculated in Fig. 7. The relationship tends to a linear one for voltages above 0.1V, which could be attributed to the ultrathin dielectric used. This will give rise to a maximum tunneling conductance per unit area of $G_g \approx 120 \, S/\mu m^2$.

The last standing question is the necessity of a symmetric junction. In fact, since the recombinations are expected to take place at the anode surface, changing the cathode material should have not any effect on the basic operation of the device, as long as fabrication is not considered as an additional challenge. However, the zero reference voltage would shift accordingly, which will be a result of net difference in the work functions of graphene and the replacing cathode material. Typical values for different cathode materials are enlisted in Table 1. However, it has to be kept in mind that asymmetric junctions with one metal electrode cause additional non-trivial shifts in calculation of $E_F$; this alteration of $E_F$ has been deeply studied using Density Functional Theory simulations [35]. A final remark may be also made on the orientation of graphene with respect to the BN layer. As it has been discussed through numerous experiments, at a precise mis-orientation angle between the graphene and BN, a mini-gap as large as 53meV may open [36-40]. This will definitely influence the long wavelength performance of the device.

### III. Conclusion

In conclusions, we investigate the possibility of light emission from graphene/dielectric sandwich structures, which radiative recombinations plus quantum tunneling will determine the optical spectrum, and therefore the color, of the devices. If the physical mechanism(s) behind the spontaneous emission in graphene-like materials is fully understood, it is hoped that these light emitting devices will open up their ways in the industry market soon afterwards.

**Acknowledgements**

The author was supported in part through a research scholarship provided by the Georgia Institute of Technology during a visiting appointment in 2011-2012.

**Appendix A**

As for the non-radiative recombinations and thermalization of carriers, a portion of the bias current in Fig. 7a should be assigned to this mechanism. We may decompose the bias current into two radiative $J_r$ and non-radiative $J_{nr}$ terms as

$$J(V) = J_r(V) + J_{nr}(V)$$

(A1)

where only the first term $J_r$ is responsible for emission of photons, giving rise to the luminance (3). All other carrier loss mechanisms outside the tunnel junction and non-radiative recombinations due to rapid thermalization of carriers may be assigned to the second term $J_{nr}$. Here, we can now define phenomenological characteristic radiative and non-radiative recombination time-constants (c.f. Appendix B [31]) as

$$J_r(V) \approx \frac{C_g V}{\tau_r}$$
$$J_{nr}(V) \approx \frac{C_g V}{\tau_{nr}}$$

(A2)

These relations should roughly hold in the linear operation regime of the device as discussed above. The radiative term is related to the luminance $L_v(\omega)$ as

$$L_v(\omega) = \frac{C_g V}{q} \hbar\omega \frac{\tau_r^{-1}}{\tau_r^{-1} + \tau_{nr}^{-1}} = \frac{C_g V}{q} \hbar\omega \frac{1}{1 + \frac{\tau_r}{\tau_{nr}}}$$

(A3)

Relating to the current density $J$ we get

$$L_v(\omega) = \frac{\hbar\omega C_g}{q G_g \left(1 + \frac{\tau_r}{\tau_{nr}}\right)} J$$

(A4)

By use of this relation and knowledge of the characteristic non-radiative recombination time $\tau_{nr}$, one may roughly estimate the radiative recombination time $\tau_r$, and thus the internal quantum efficiency $\eta \cong 1/\left(1 + \frac{\tau_r}{\tau_{nr}}\right)$.

**Appendix B**

In this Appendix, we demonstrate the validity of relation (8). From relation (14) of the paper [28] we have

$$\hat{e} \cdot \langle c\mathbf{k}|\mathbf{v}|v\mathbf{k}\rangle = \pm \frac{j}{\hbar} a\gamma_0 \frac{\sqrt{3}}{2} \frac{e_y \Delta k_x - e_x \Delta k_y}{\sqrt{\Delta k_x^2 + \Delta k_y^2}} = \pm j v_F \hat{e} \cdot \hat{z} \times \frac{\Delta \mathbf{k}}{|\Delta \mathbf{k}|}$$

(B1)

where $v_F = \sqrt{3} a\gamma_0/2\hbar$ is the Fermi velocity, $\hat{z}$ is the unit vector along z-axis, and $\hat{e}$ is the in-plane unit polarization vector of electric field. As a result, we may write down

$$\langle c\mathbf{k}|\mathbf{v}|v\mathbf{k}\rangle = j v_F (\hat{z} \times \hat{\kappa})$$

(B2)

in which $\hat{\kappa} = +(\mathbf{k} - \mathbf{K})/|\mathbf{k} - \mathbf{K}|$ or $\hat{\kappa} = -(\mathbf{k} - \mathbf{K}')/|\mathbf{k} - \mathbf{K}'|$ depending on the choice of Dirac point. Therefore magnitude of the transition element vector (B2) is a simple constant

$$|\langle c\mathbf{k}|\mathbf{v}|v\mathbf{k}\rangle| = v_F$$

(B3)

Consequently, the emission of single-photons is associated with a constant transfer element since the polarization vector is circular, that is $\hat{e} = \frac{1}{\sqrt{2}}(\hat{x} \pm j\hat{y})$. Hence, we first get

$$(\hat{e} \cdot \langle c\mathbf{k}|\mathbf{v}|v\mathbf{k}\rangle)^2 = v_F^2 \tag{B4}$$

which is independent of photon energy. The second result comes from the electric dipoles of the transitions in Graphene. We have

$$\langle c\mathbf{k}|\mathbf{v}|v\mathbf{k}\rangle = \frac{j}{\hbar}\langle c\mathbf{k}|[\mathbb{H},\mathbf{r}]|v\mathbf{k}\rangle = \frac{j}{\hbar}[\langle c\mathbf{k}|\mathbb{H}\mathbf{r}|v\mathbf{k}\rangle - \langle c\mathbf{k}|\mathbf{r}\mathbb{H}|v\mathbf{k}\rangle] = \frac{j}{\hbar}[E_c(\mathbf{k}) - E_v(\mathbf{k})]\langle c\mathbf{k}|\mathbf{r}|v\mathbf{k}\rangle \tag{B5}$$

For a resonant transition, we have $E_c(\mathbf{k}) - E_v(\mathbf{k}) = \hbar\omega$, and therefore from (B2) we get

$$\langle c\mathbf{k}|q\mathbf{r}|v\mathbf{k}\rangle = \frac{qv_F}{\omega}(\hat{z} \times \hat{\kappa}) \tag{B6}$$

or

$$\rho = \frac{qv_F}{\omega} \tag{B7}$$

in which $\rho = |\langle q\mathbf{r}\rangle|$ is the magnitude of the electric dipole moment. Expressing in standard Debye units, we have

$$\rho = 10^{21}\frac{qv_F}{2\pi}\lambda \text{ (Debye)} \tag{B8}$$

where $\lambda$ is the wavelength and all parameters on the right-hand-side are given in SI units. This is the same as relation (8).